\documentclass[journal]{IEEEtran}
\usepackage{graphicx}
\usepackage{amsmath}
\usepackage{amssymb}
\usepackage{amsbsy}
\usepackage{multirow}

\begin{document}
\title{Maximum Phase Modeling for Sparse Linear Prediction of Speech}
%

%\author{Thomas Drugman,~\IEEEmembership{Member,~IEEE,}      
\author{Thomas Drugman, \textit{Member, IEEE}}      
%\thanks{T. Drugman is with the TCTS Lab, University of Mons, and is supported by FNRS.}

\markboth{IEEE Signal Processing Letters}%
{Shell \MakeLowercase{\textit{et al.}}: Bare Demo of IEEEtran.cls for Journals}

\maketitle

\begin{abstract}
%\boldmath

Linear prediction (LP) is an ubiquitous analysis method in speech processing. Various studies have focused on sparse LP algorithms by introducing sparsity constraints into the LP framework. Sparse LP has been shown to be effective in several issues related to speech modeling and coding. However, all existing approaches assume the speech signal to be minimum-phase. Because speech is known to be mixed-phase, the resulting residual signal contains a persistent maximum-phase component. The aim of this paper is to propose a novel technique which incorporates a modeling of the maximum-phase contribution of speech, and can be applied to any filter representation. The proposed method is shown to significantly increase the sparsity of the LP residual signal and to be effective in two illustrative applications: speech polarity detection and excitation modeling.

\end{abstract}


% \fi
%\IEEEpeerreviewmaketitle

\let\thefootnote\relax\footnotetext{
\\T. Drugman is supported by FNRS, and with TCTS Lab, University of Mons, Belgium. \textit{Address:} 31 Boulevard Dolez, 7000 Mons, Belgium. \emph{Phone:} +3265374749. \emph{Email:} thomas.drugman@umons.ac.be.}

\section{Introduction}
\label{sec:Intro}

Linear prediction (LP) is an omnipresent analysis technique in speech processing. It has been successfully applied in several voice technology applications such as speech coding, synthesis, analysis or recognition \cite{Quatieri}. LP analysis relies on a source-filter model in which the speech signal is obtained by passing an excitation through an all-pole filter. In the traditional LP analysis approach, the prediction coefficients are determined such that the $l_2$-norm of the residual signal (i.e. the difference between the observed and predicted signals) is minimized \cite{Makhoul}. This is known to work rather well for unvoiced sounds where the excitation signal can be assumed to be Gaussian and independently and identically distributed \cite{Makhoul}. Nonetheless, for voiced sounds where the excitation signal exhibits quasi-periodic strong peaks, this assumption does not hold. In this case, the hypothesized excitation source is a quasi-periodic pulse train. Minimizing the variance of the residual signal then turns out to not be an appropriate criterion, as this approach is known to suffer from problems such as overemphasis on peaks and cancellation of errors \cite{Makhoul}.

A better criterion would be to maximize the sparsity of the residual signal. In that perspective, its $l_0$-norm should be ideally minimized. This however yields a combinatorial optimization problem. Instead, various approaches \cite{Giacobello} have investigated the minimization of the $l_1$-norm which is a convex relaxation of the $l_0$-norm problem and which can be solved using convex programming methods.

Across all aforementioned techniques, poles are expected to lie within the unit circle in the z-plane, otherwise the filter is considered to be unstable \cite{Knockaert}, \cite{Giacobello}. In the $l_2$-norm problem, this is guaranteed but is however not true with the $l_1$-norm \cite{Knockaert}. Generally, when poles are found outside the unit circle, pole reflection is applied \cite{Giacobello}. However, during the production of voiced sounds, the glottal flow is known to exhibit a maximum-phase (i.e. anticausal) component \cite{Quatieri}, \cite{MixedPhase}, which is therefore generally not modeled in conventional LP analysis.

%In \cite{Giacobello}, authors report that the percentage of unstable filters is low (around $2\%$) with a maximum radius for a pole of $\rho_{max} = 1.0259$. 

The goal of this paper is to propose a solution to incoporate a maximum-phase modeling in the LP analysis of speech. The proposed method can be applied to any LP-based method and is shown to significantly improve the sparsity of the residual signal. The paper is structured as follows. Section \ref{sec:LP} first establishes the fundamentals of linear prediction. The existence of a maximum-phase component in speech is explained in Section \ref{sec:MP}, where the motivations of this work are given. The proposed method is described in Section \ref{sec:Method}. Its efficiency is then confirmed in Section \ref{sec:Exp} through a comprehensive evaluation. Section \ref{sec:Conclu} finally concludes the paper.

%\vspace{-4pt}

%%%%%%%%%%%%%%%%%%%%%%%%%%%%%%%%%%%%%%%%%%%%%%%
\section{Linear Prediction: Problem Formulation}
\label{sec:LP}

The auto-regressive (AR) model of speech assumes that a speech sample $s(n)$ can be written as a linear combination of its $K$ past samples: $s(n) = \sum_{k=1}^{K} a_k s(n-k) + r(n)$, where $K$ is the prediction order, $a_k$ are the prediction coefficients and $r(n)$ is the prediction error, also called residual signal or residue. Based on the observation of a sequence of speech samples, the optimization problem aims to find an estimate of the prediction coefficient vector $\mathbf{\hat{a}} \in \mathbb{R}^{K}$ such that the prediction error is minimized. The LP analysis problem can then be written as the minimization of the $l_p$-norm of the residual signal.

%\begin{equation}
%s(n) = \sum_{k=1}^{K} a_k s(n-k) + r(n) 
%\label{eq:LP}
%\end{equation}

%\begin{equation}
%\mathbf{\hat{a}} = \underset{\mathbf{a}}{\arg\min} \parallel \mathbf{r} \parallel _{p}^{p}, \mathbf{r} = \mathbf{s} - \mathbf{Sa}, 
%\label{eq:matrix}
%\end{equation}

The conventional approach considers the minimization of the $l_2$-norm problem which can be solved in a rapid way by exploiting the Toeplitz structure of the correlation matrix, as in the widely-used Levinson-Durbin algorithm \cite{Levinson}. Nonetheless, it is known that the $l_2$-norm criterion is highly sensitive to outliers. As a consequence, this approach will favor solutions with many small non-zero values rather than a sparse solution containing a limited number of non-zero values \cite{Giacobello}. Sparse solutions should be preferred as the target excitation source in voiced sounds is expected to be a quasi-period impulse train.

The ideal solution maximizing sparsity involves the minimization of the $l_0$-norm of the residual signal. Unfortunately this leads to a combinatorial problem which cannot be solved in polynomial time. Instead, several studies have addressed solving the $l_1$-norm, moving closer to the original $l_0$-norm problem \cite{Giacobello}. This is possible thanks to the improvements in convex optimization algorithms (e.g. using interior point methods \cite{Boyd}). Other approaches have proposed a weighted $l_2$-norm LP analysis in which a weighting function is used to give less emphasis to the samples around a strong excitation \cite{WLP,Alku-WLP,Vahid}. Compared to the conventional $l_2$-norm, aforementioned methods have proved to be more efficient for coding and to provide better estimates of the spectral envelope.

%Recently, the technique described in \cite{Vahid} proposes to use a weighted $l_2$-norm where the weighting function gives less emphasis to the samples around the detected glottal closure instants (GCIs). GCIs are the instants of significant excitation of the vocal tract during which the prediction error is expected to be large \cite{SEDREAMS}. The resulting method benefits from the computational complexity of the $l_2$-norm problem and is shown to achieve sparsity results comparable to what is obtained using the $l_1$-norm.

%%%%%%%%%%%%%%%%%%%%%%%%%%%%%%%%%%%%%%%%%%%%%%%
\section{Motivation of this Work}
\label{sec:MP}
According to the mixed-phase model \cite{MixedPhase}, speech is composed of both minimum-phase (i.e causal) and maximum-phase (i.e anticausal) components. While the vocal tract impulse response and the \emph{return phase} of the glottal component are minimum-phase signals, the \emph{open phase} of the glottal flow is known to be maximum-phase \cite{MixedPhase}. In \cite{Gardner}, it was proved that the use of an anticausal all-pole filter for the glottal pulse is necessary to resolve magnitude and phase information correctly. Unfortunately, deconvolving the minimum and maximum-phase components of speech is a complex problem which suffers from robustness issues \cite{Drugman-Robustness}, \cite{Chirp}.

\begin{figure}[ht]
\centering
\includegraphics[width=0.4\textwidth]{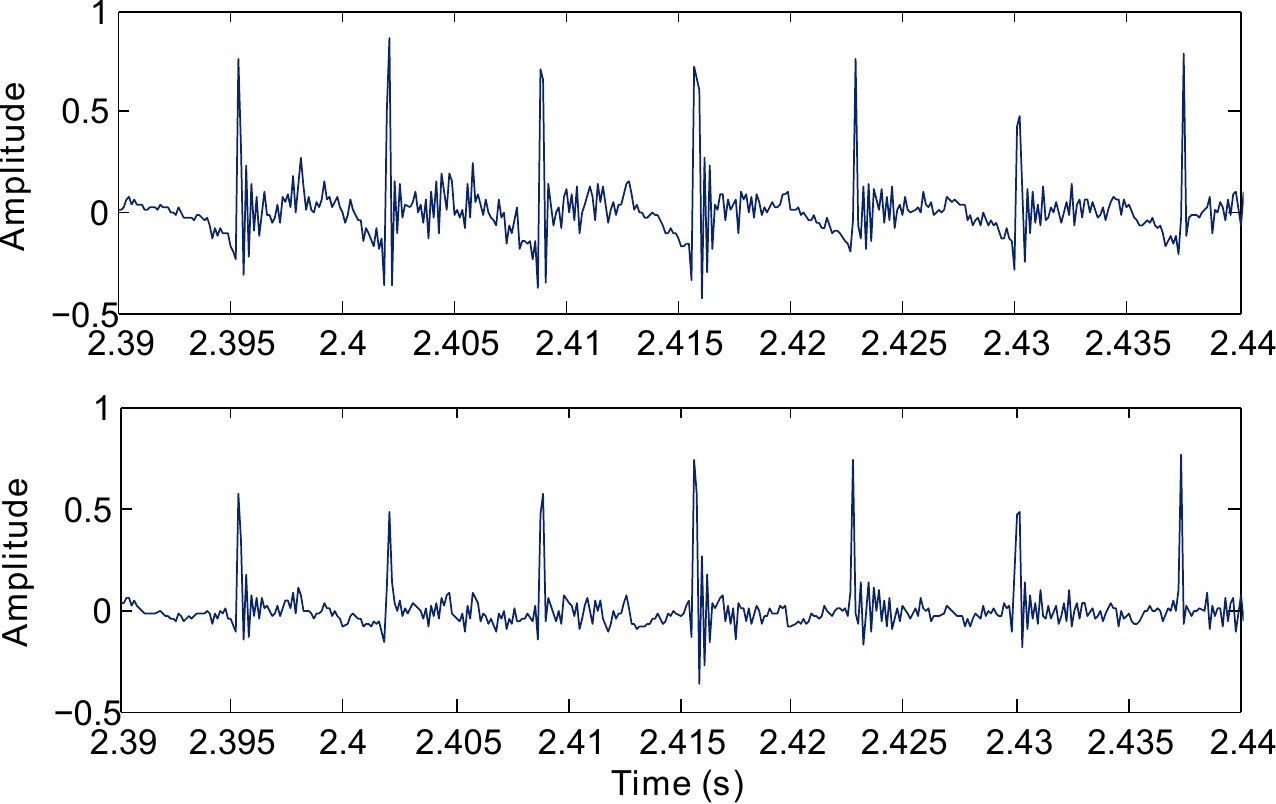}
\caption{Example of residual signals obtained using: \emph{top panel:} the conventional $l_2$-norm LP analysis; \emph{bottom panel:} the proposed $l_2$-norm LP analysis.}  
\label{fig:BasicExample}
\vspace{-8pt}
\end{figure}

Despite the mixed-phase nature of speech, existing LP analysis approaches generally apply pole reflection when poles are found outside the unit circle. The resulting residue consequently contains a maximum-phase component which has not been captured in the AR modeling. The top plot of Fig. \ref{fig:BasicExample} shows an example of residue obtained using the conventional $l_2$-norm LP analysis. Glottal Closure Instants (GCIs, \cite{GCI}) appear as strong quasi-periodic discontinuities in this signal. It can be noticed that the segments preceding GCIs follow a systematic shape with clearly non-zero valued samples. This is because the residue exhibits a persistent maximum-phase component which has not been completely removed  by the standard LP analysis. An intuitive way to explain this would be to reverse the time axis: it can be understood that this remaining component could be modeled by a simple AR filter with a limited number of poles. The operation of reversing the time axis is equivalent to inversing the causality of the signal. As a consequence, modeling the maximum-phase component of speech could potentially engender sparser residual signals. This is the precise goal of the proposed method. The bottom plot of Fig. \ref{fig:BasicExample} shows the residue when applying the proposed technique to the $l_2$-norm LP analysis. It can be observed that in this case the maximum-phase component has been removed almost completely.

%%%%%%%%%%%%%%%%%%%%%%%%%%%%%%%%%%%%%%%%%%%%%%%
\section{Proposed Method}
\label{sec:Method}

\subsection{Description}
\label{ssec:Description}

% \emph{http://tcts.fpms.ac.be/\~drugman/MaxP.zip}

\begin{figure}[ht]
\centering
\includegraphics[width=0.45\textwidth]{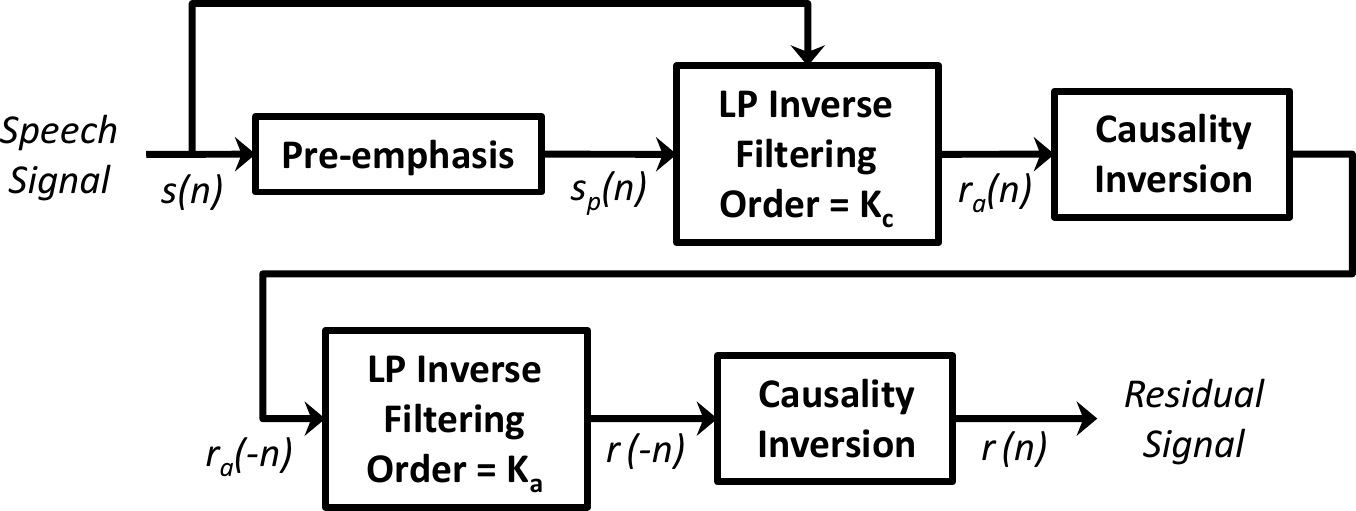}
\caption{Workflow of the proposed method.}  
\label{fig:Workflow}
\vspace{-4pt}
\end{figure}

The workflow of the proposed method is given in Fig. \ref{fig:Workflow}. Note that a Matlab implementation of this technique can be found at \emph{tcts.fpms.ac.be/$\sim$drugman/Toolbox}. The speech signal is first standardly framed using overlapping windows. Because mixed-phase decomposition is a challenging problem suffering from robustness issues in realistic recording conditions \cite{Drugman-Robustness}, the proposed approach circumvents this hindrance by relying on steps of preemphasis and causality inversion. The residue is obtained by two succesive inverse filtering operations: in the first one, we aim at removing the minimum-phase contribution, while the second targets removing the maximum-phase component. The coefficients of the 2 filters are estimated by LP analysis, as explained below.

The minimum-phase component of speech is mostly related to the vocal tract \cite{MixedPhase}, whose dominant poles are due to the first formant $F1$. On the opposite, the maximum-phase contribution is due to the glottal open phase, which is characterized by the \emph{glottal formant} $Fg$, whose range is known to cover $[F_0,3*F_0]$ \cite{GlottalFormant}. The goal of the first LP analysis (of order $K_c$) is to model the minimum-phase component. In order to minimize the effect of the anticausal contribution, and maximize the impact of the causal component, the coefficients of the first LP analysis are estimated on a pre-emphasized version $s_p(n)$ of the speech signal $s(n)$. Pre-emphasis is conventionnaly achieved by using a single real zero in $\alpha$. Preemphasis therefore balances the energy in the speech spectrum such that low frequencies do not dominate during the first LP analysis. The choice of $\alpha$ used for pre-emphasis will then result from a trade-off, as it will be discussed in Section \ref{ssec:Parameter}.

After the first LP inverse filtering, the resulting signal $r_a(n)$ is mostly dominated by the anticausal component of speech, as the causal contribution has been substantially removed. The second LP analysis precisely aims at modeling this maximum-phase component. To force this, we apply causality inversion by reversing the time axis. After causality inversion, the original anticausal component is now seen as causal, and can be modeled by a standard LP analysis of order $K_a$. The signal $r_a(-n)$ is then inverse filtered to get $r(-n)$, and the final residual signal $r(n)$ is simply obtained by reversing the time axis in its original direction.

The success of the proposed method lies in two key concepts: \emph{i)} since conventional LP analysis assumes the signal to be minimum-phase, reversing causality is a solution to force the modeling of the maximum-phase component of speech; \emph{ii)} preemphasis is essential as it guides the two successive LP analyses; its goal is to minimize the effects of the maximum-phase contribution, such that the first LP analysis is mostly driven by the minimum-phase component of speech. An advantage of the proposed method is that it can be applied to any LP analysis technique, and even transposed to other filter representations such as MFCCs or Mel-Generalized Cepstral (MGC) coefficients \cite{MGC}. The experiments led in this paper however only focus on its usefulness for LP analysis.

\subsection{Parameter Settings}
\label{ssec:Parameter}
The proposed method makes use of 3 parameters: the total prediction order $K = K_c + K_a$, the prediction order $K_a$ for the maximum-phase modeling, and the preemphasis coefficient $\alpha$. $K$ can be fixed as done by standard LP methods. The influence of the two other parameters on the engedered sparsity is now studied on a development set containing 1000 sentences from the TIMIT corpus \cite{TIMIT} (balanced across genders). As sparsity metrics, we use the Gini index \cite{Gini} as it was the only sparsity metrics in \cite{Kurtosis} to meet the six attributes one can expect from a sparsity measure. Higher values of the Gini index imply a higher level of sparsity. Throughout our development experiments, we observed that the Gini index of the residual signal $r(n)$ reaches higher values when $K_a$ is fixed to 2 or 3. This goes in line with the speech production model according to which the maximum-phase component is due to the glottal open phase which can be modeled by two anticaussal poles \cite{MixedPhase}. $K_a$ is therefore fixed to 2 in the remainder of this paper. 

The setting of $\alpha$ is linked with the fact that minimum and maximum-phase components are assumed to affect different spectral bands, and results from a tradeoff. It must be such that, during the first LP analysis: \emph{i)} the minimum-phase component (whose dominant poles are due to $F1$) will be properly modeled; \emph{ii)} the effects of the first harmonics and of the maximum-phase component (dominated by $Fg$) are minimized. $\alpha$ is then expected to be dependent in a certain extent upon $F_0$, and consequently upon the speaker gender.

%Preemphasis is fundamental in the proposed technique and its usefulness is linked with the fact that minimum and maximum-phase components are assumed to affect different spectral bands. The setting of $\alpha$ then results from a trade-off. It must be such that, during the LP analysis of $s_p(n)$: \emph{i)} the minimum-phase component (whose dominant poles are due to the first formant $F1$) will be properly modeled; \emph{ii)} the effects of the first harmonics and of the maximum-phase component (dominated by the glottal formant whose range is $[F_0,3*F_0]$ \cite{GlottalFormant}) are minimized. $\alpha$ is then expected to be dependent upon $F_0$, and consequently upon the speaker gender.

%Fig. \ref{fig:AlphaSetting} shows the sensitivity of the proposed approach to $\alpha$ for both male (left column) and female (right column) speakers. The top plots display the evolution of the kurtosis as a function of $\alpha$ (here considered to be constant in a given sentence) for 3 techniques: the conventional $l_2$-norm LP analysis (\emph{LP2}), the same technique when applying the proposed approach (\emph{Max-LP2}) and when the same process is applied without any causality inversion (\emph{Min-LP2}). It can be noticed that if $\alpha$ is properly chosen, the gain in sparsity over both \emph{LP2} and \emph{Min-LP2} can be dramatic, especially for female voices. Nonetheless, this choice turns out to be crucial for male speakers where a degradation in sparsity is possible.

\begin{figure}[ht]
\centering
\includegraphics[width=0.45\textwidth]{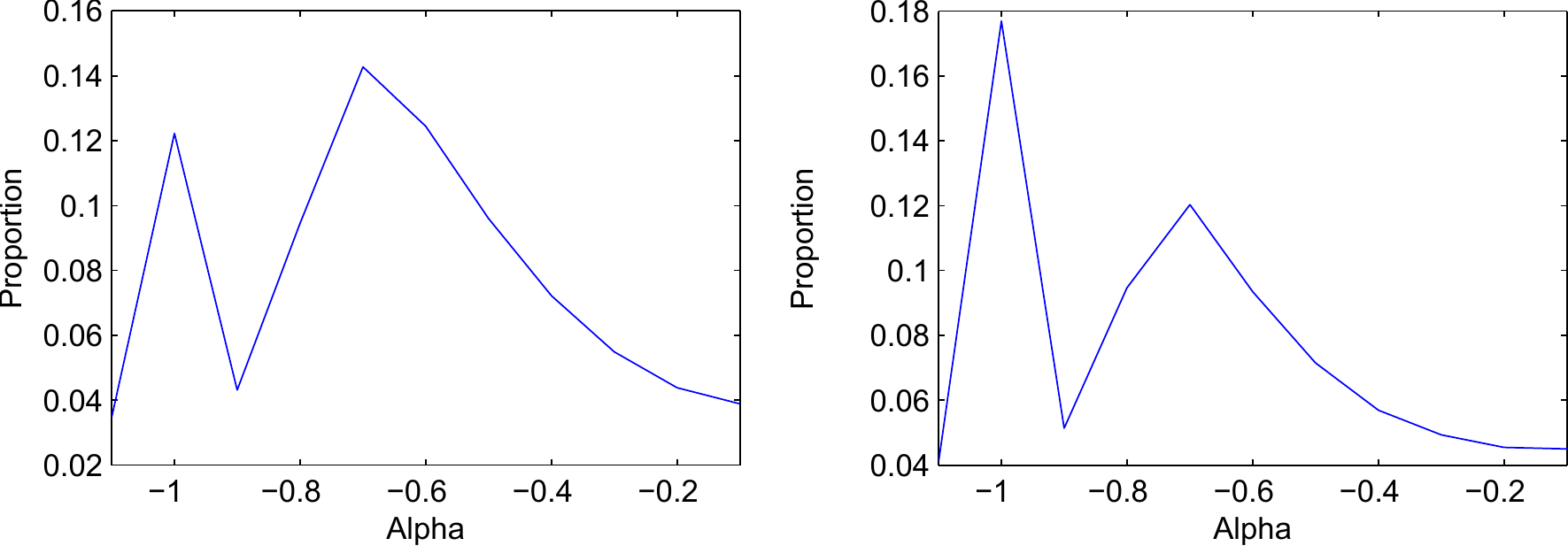}
\caption{Distribution of the optimal coefficient $\alpha$ for male (left panel) and female (right panel) speakers.}
\label{fig:AlphaSetting}
\vspace{-8pt}
\end{figure}

Fig. \ref{fig:AlphaSetting} exhibits the distribution of the optimal $\alpha$ values across the development set. For both genders, the distribution is bimodal with two clear modes in $\alpha = -1$ and in $\alpha = -0.7$ (this latter being more spread). As a consequence, the setting of $\alpha$ is not univocal. The approach we adopt in the rest of this paper is then to investigate, for each frame to be analyzed, two possible values for $\alpha$ (-1 and -0.7) and ultimately select the one engendering the greater sparsity, i.e. maximizing the Gini index of the resulting residue $r(n)$. 

%The computational load of the proposed method is then roughly the double than that of a standard LP analysis.

%Two representative techniques for the proposed approach are considered in the following: \emph{MaxF-LP} which is based on the "full" range for the $\alpha$ search [-1.2,0] by step of 0.1, and \emph{MaxR-LP} which investigates only the two modal values $-1$ and $-0.3$. Roughly speaking, \emph{MaxF-LP} induces an increase of the computational complexity by a factor 13, while it is doubled for \emph{MaxR-LP}.

%\caption{Influence of the preemphasis coefficient $\alpha$ for male (left column) and female (right column) speakers. The top plot indicates the kurtosis using: the proposed approach applied to the standard $l_2$-norm LP analysis (\emph{Max-LP2}), the proposed workflow where the 2 operations of causality inversion are omitted (i.e. only the minimum-phase is modeled - \emph{Min-LP2}), the conventional $l_2$-norm LP analysis (\emph{LP2}). The bottom panels show the distribution of the optimal $\alpha$ across the sentences included in the development set.}

%%%%%%%%%%%%%%%%%%%%%%%%%%%%%%%%%%%%%%%%%%%%%%%
\section{Experiments}
\label{sec:Exp}

Our experiments are divided into three parts. Section \ref{ssec:Sparsity} investigates the sparsity and computational complexity engendered by the proposed technique. Sections \ref{ssec:Polarity} and \ref{ssec:Excitation} then address its efficiency in two illustrative applications: speech polarity detection and excitation modeling. Note that a common way to assess the efficiency of a LP analysis technique is to compute the spectral distortion (SD) between a reference envelope and the estimated predictive model \cite{Giacobello}. This would be meaningless here since the improvement brought by the proposed method relies on the exploitation of phase properties. SD calculation indeed only involves the amplitude component of the Fourier spectrum, and discards its phase counterpart. 

\subsection{Sparsity and computational complexity}
\label{ssec:Sparsity}

The proposed approach is here applied to 3 techniques: the conventional $l_2$-norm (\emph{LP2}), the weigthed $l_2$-norm (\emph{WLP2}) proposed in \cite{Vahid} and the $l_1$-norm (\emph{LP1}) LP analyses. \emph{WLP2} applies a weighting function to give less emphasis to the samples around GCIs \cite{Vahid}. GCIs are here determined using the SRH (for F0 tracking \cite{SRH}) and SEDREAMS algorithms \cite{SEDREAMS}. For the minimization of the $l1$-norm, we use the publicly available l1-magic toolbox \cite{Magic} based on a primal-dual interior points optimization \cite{Boyd}. In our experiments, we compare the conventional implementation of these 3 techniques to their declined version based on the proposed MaxP (standing for maximum-phase) method. Across all techniques, framing is achieved by applying a 25ms-long Hanning window shifted every 5 ms. The 3000 longest sentences of the TIMIT corpus \cite{TIMIT} (balanced across genders and not included in the development set) are used for the evaluation. As in \cite{Vahid}, sentences are resampled at 8 kHz and $K$ is fixed to 13.

Three sparsity metrics are here used to assess the performance of the LP techniques: the kurtosis, Hoyer measure \cite{Hoyer} (which is a normalized version of the $\frac{l_2}{l_1}$ measure) and the Gini index \cite{Gini} of the residual signal. These 3 metrics were shown in \cite{Kurtosis} to be the 3 most appropriate measures to reflect the sparsity of a signal, as they respectively meet 3, 5 and 6 out of the 6 essential attributes of a sparsity metrics. More precisely, we consider in the following the relative improvement in sparsity of the residue over the speech signal. In other words, considering a given sparsity measure $SM(x)$ (which can be any of the three aforementioned metrics), we evaluate the sparsity improvement as: $\frac{SM(r(n))-SM(s(n))}{SM(s(n))}$, where $s(n)$ and $r(n)$ are the original speech signal and its residue.

\begin{table}[!ht]
\centering
\begin{tabular}{| c | c | c | c | c |}
\textbf{Metrics} & \textbf{Method} & \textbf{LP2} & \textbf{WLP2} & \textbf{LP1}\\
\hline    
\hline    
\multirow{2}{*}{Kurtosis} & Conventional & 250 & 403 & 392\\  
 & MaxP & 364 & 432 & 468\\
\hline
\hline    
\multirow{2}{*}{Hoyer} & Conventional & 17.6 & 26.0 & 29.4\\  
 & MaxP & 24.0 & 29.3 & 30.5\\    
\hline
\hline    
\multirow{2}{*}{Gini} & Conventional & 3.48 & 4.71 & 7.54\\  
 & MaxP & 5.64 & 6.82 & 7.36\\    
\hline    
\end{tabular}
\caption{Relative improvement in sparsity (in \%) over the speech signal, using the compared methods.}
\label{tab:Sparsity}
\vspace{-16pt}
\end{table}

Results are summarized in Table \ref{tab:Sparsity}. It can be observed that the proposed MaxP method leads to a considerable increase of sparsity. This was reflected across 8 out of the 9 configurations (3 techniques and 3 metrics). However, for the unfavorable case (using the LP1-based techniques and the Gini index), a paired t-test revealed no significant differences, while all other results showed a statistically highly significant improvement ($p<0.001$). Finally it is worth noting that we did not observe any gender dependency through our experiments.

%\subsection{Computational complexity}
%\label{ssec:Computation}

The computational complexity of the methods is now assessed by the Relative Computation Time (RCT), defined as the ratio between the computation time over the sound duration. Table \ref{tab:RCT} shows averaged RCTs obtained for our Matlab implementations and with a Intel Core i7 3.0 GHz CPU with 16GB of RAM. The proposed MaxP method results in an increase of complexity by a factor varying between 2.8 and 3.9. This is because 2 LP analyses of order $K_a$ and 2 of order $K_c$ are achieved, instead of only one of order $K$. Note also that WLP2-based techniques require in addition the estimation of GCIs, which is performed in a RCT of 7.3\%.

\begin{table}[!ht]
\centering
\begin{tabular}{| c | c | c | c | c | c | c | c |}
\hline 
\textbf{LP2} & \textbf{MaxP-LP2} & \textbf{WLP2} & \textbf{MaxP-WLP2} & \textbf{LP1} & \textbf{MaxP-LP1}\\
\hline    
1.6 & 4.5 & 2.2 & 7.4 & 41.2 & 160\\
\hline    
\end{tabular}
\caption{Relative computation time (in \%) for the compared methods.}
\label{tab:RCT}
\vspace{-32pt}
\end{table}

%
%\begin{table}[!ht]
%\centering
%\begin{tabular}{| c | c | c |}
%\textbf{Method} & \textbf{Male} & \textbf{Female} \\
%\hline    
%\hline    
%LP2 & 91.2 & 48.5 \\
%\hline    
%MaxR-LP2 & 134.6 & 90.0 \\
%\hline    
%MaxF-LP2 & 130.5 & 95.4 \\
%\hline
%\hline
%WLP2 & 120.6 & 77.3 \\
%\hline    
%MaxR-WLP2 & 165.9 & 113.1 \\
%\hline    
%MaxF-WLP2 & 165 & 112.4 \\
%\hline
%\hline
%LP1 & 116.7 & 77.5 \\
%\hline    
%MaxR-LP1 & 168.8 & 112.2 \\
%\hline    
%\end{tabular}
%\caption{Sparsity results for male and female voices.}
%\label{tab:Sparsity}
%\end{table}

\subsection{Application to Speech Polarity Detection}
\label{ssec:Polarity}
The origin of a polarity in the speech signal stems from the asymmetric glottal waveform exciting the vocal tract resonances. Detecting the speech polarity is required in various applications such as concatenative synthesis, glottal source processing or in the great majority of pitch-synchronous techniques. In \cite{RESKEW}, the Residual Excitation Skewness (RESKEW) approach has been proposed to automatically determine the speech polarity. RESKEW exploits the statistical skewness of two excitation signals: the LP residual, and a rough approximation of the glottal source. Since the skewness is known to be a measure of the asymmetry of a probability density function, it is used here as an estimator of the asymmetry of the glottal excitation. As the LP residue and the glottal source are known to have an opposite polarity, the sign of their differenced skewness indicates the speech polarity \cite{RESKEW}.

\begin{figure}[ht]
\centering
\includegraphics[width=0.4\textwidth]{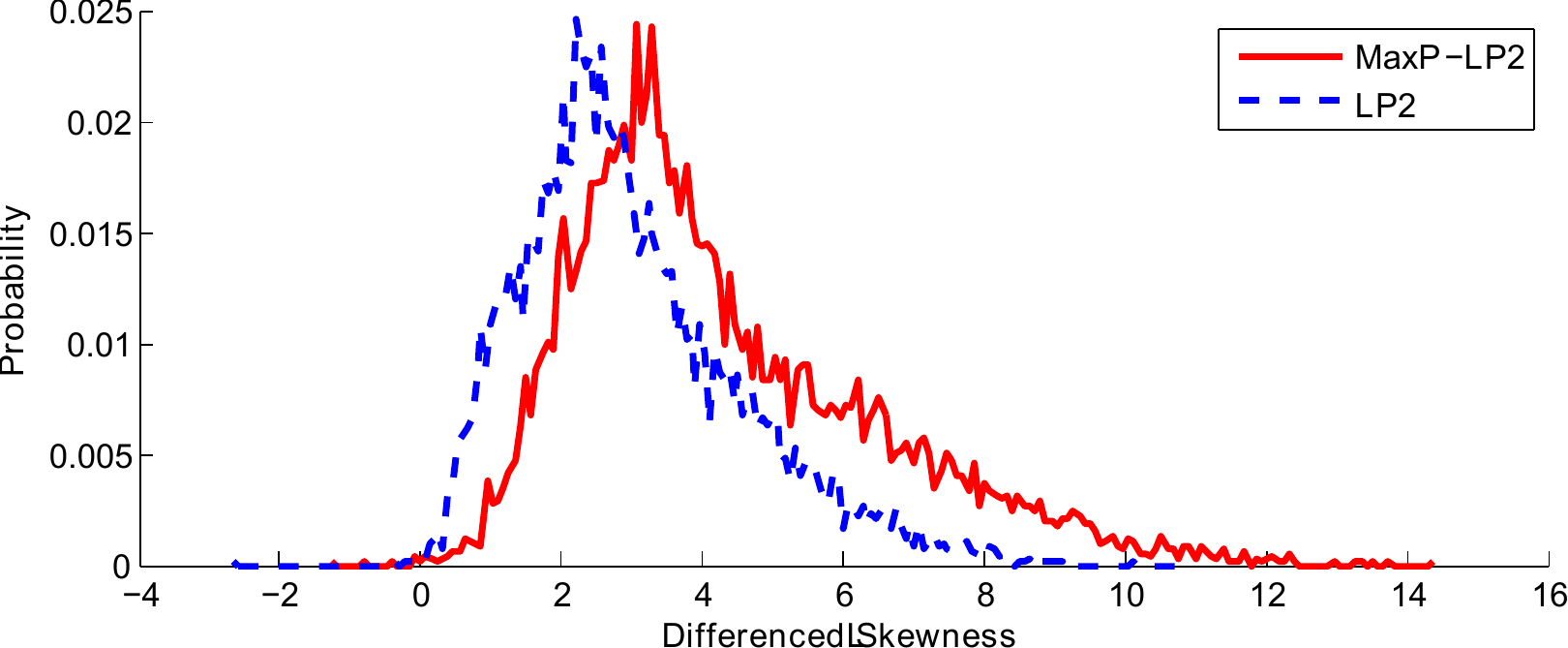}
\caption{Distribution of the differenced skewness used in the RESKEW method to determine the speech polarity.}  
\label{fig:Polarity}
\vspace{-16pt}
\end{figure}

Fig. \ref{fig:Polarity} displays the distribution of the differenced skewness using either the traditional \emph{LP2} or the proposed \emph{MaxP-LP2} residual signal. This is done for the 10 corpora considered in \cite{RESKEW}, covering in total 7.5 hours of speech. Note that we here used \emph{MaxP-LP2} for the fact that it does not require GCIs. This is because polarity detection is the very first step in an analysis workflow and should then be as simple as possible. The advantage of using \emph{MaxP-LP2} is clearly noticed in Fig. \ref{fig:Polarity}. The distribution of the differenced skewness is indeed observed to move off from zero, reducing therefore considerably the risk of confusion in the polarity determination.
% Only 3 files among the 8545 contained in the 10 corpora were erroneously detected with the \emph{MaxR-LP2} approach (against 5 for the conventional \emph{LP2} technique).

\subsection{Application to Excitation Modeling}
\label{ssec:Excitation}
The usefulness of the proposed technique will now be studied in the context of excitation modeling. In \cite{DSM}, the Deterministic plus Stochastic Model (DSM) of the residual signal was proposed. DSM consists of two contributions acting in two distinct spectral bands delimited by a maximum voiced frequency. The deterministic part models the periodicity in the low frequencies, while the stochastic component is a time-modulated high-frequency noise accounting for the glottal turbulences. Both components are extracted from an analysis performed on a dataset of GCI-synchronous windowed residual frames. The deterministic component arises from an orthonormal decomposition led on this dataset, which is achieved by Principal Component Analysis (PCA, \cite{PCA}). It has been shown in \cite{DSM} that the resulting first eigenvector can be assumed to model the deterministic component of the residue.

\begin{figure}[ht]
\centering
\includegraphics[width=0.4\textwidth]{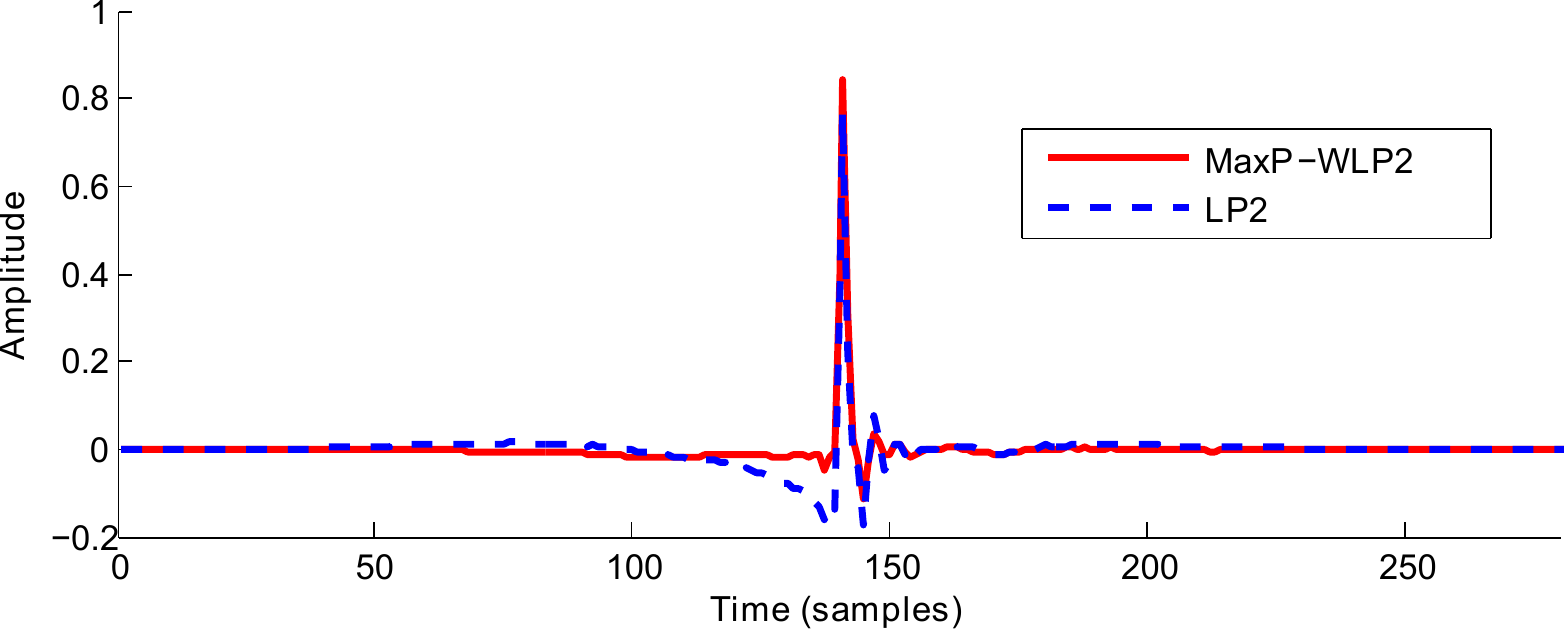}
\caption{First eigenvector for the male speaker AWB.}  
\label{fig:DSM}
\vspace{-16pt}
\end{figure}

Fig. \ref{fig:DSM} shows the first eigenvector for the male speaker AWB from the CMU-ARTIC database \cite{ARCTIC}, using the \emph{LP2} and \emph{MaxP-WLP2} techniques. \emph{WLP2} is here used since GCIs are already necessary for DSM to perform a pitch-synchronous analysis. In the case of \emph{LP2}, the first eigenvector exhibits a waveform at the left of the GCI which is similar to what is described by models of the glottal source \cite{GFmodels}. This is because the \emph{LP2} residue exhibits a persistent maximum-phase component which has not been eliminated. As a consequence, this component is reflected in the open phase of the first eigenvector. Contrastingly, the first eigenvector obtained with \emph{MaxP-WLP2} is very close to a Dirac pulse and its open phase is almost completely flat. When inspecting the eigenvalues, we observed that \emph{MaxP-WLP2} allows to cover a comparable dispersion with a reduced number of eigenvectors, which makes it interesting for speech coding.

%A small ripple, comparable to the one in the \emph{LP2} example, is noticed on the right of the GCI. This is due to the weak imperfections of the auto-regressive modeling to capture the minimum-phase component of speech. 

%%%%%%%%%%%%%%%%%%%%%%%%%%
%%%%%%%%%%%%%%%%%%%%%%%%%%

\section{Conclusion}
\label{sec:Conclu}

The goal of this paper was to propose a novel approach to the problem of sparse LP analysis. The proposed method aims to integrate a modeling of the maximum-phase component of speech, which is discarded in existing LP-based techniques. It has also the advantage of being applicable to any filter representation. The resulting technique was shown to dramatically increase the level of sparsity, independently of the LP method it is applied to. This was achieved with a minor augmentation of the computational complexity. Finally, its potentiality was confirmed in two illustrative applications: polarity detection and excitation modeling.

%The success of the proposed method lies in two key concepts: \emph{i)} preemphasis before the first LP analysis is essential as it guides the preponderance of the minimum and maximum-phase components in the two successive LP analyses; \emph{ii)} since conventional LP analysis assumes the signal to be minimum-phase, reversing causality is a solution to force the modeling of the maximum-phase component of speech.

%\section*{Acknowledgment}

\ifCLASSOPTIONcaptionsoff
  \newpage
\fi


\begin{thebibliography}{1}

\bibitem {Quatieri}
T. Quatieri: \emph{Discrete-time speech signal processing}, Prentice Hall, 2002.

\bibitem {Makhoul}
J. Makhoul: \emph{Linear Prediction: A Tutorial Review}, Proc. IEEE, vol. 63, no. 4, pp. 561--580, April 1975.

\bibitem {Giacobello}
D. Giacobello, M. Christensen, M. Murthi, S. Jensen, M. Moonen: \emph{Sparse Linear Prediction and Its Applications to Speech Processing}, IEEE Trans. on Audio, Speech and Language Processing, vol. 20, Issue 5, pp. 1644--1657, 2012.

\bibitem {Knockaert}
L. Knockaert: \emph{Stability of linear predictors and numerical range of shift operators in normed spaces}, IEEE Trans. on Inf. Theory, vol. 38, no. 5, pp. 1483--1486, 1992.

\bibitem {MixedPhase}
T.Drugman, B.Bozkurt, T.Dutoit: \emph{Causal-anticausal Decomposition of Speech using Complex Cepstrum for Glottal Source Estimation}, Speech Communication, vol. 53, Issue 6, pp. 855--866, 2011.

\bibitem {Levinson}
G. Cybenko: \emph{The Numerical Stability of the Levinson-Durbin Algorithm for Toeplitz Systems of Equations}, SIAM J. Sci. and Stat. Comput., vol. 1, Issue 3, pp. 303--319, 1980.

\bibitem {Boyd}
S. Boyd, L. Vandenberghe: \emph{Convex Optimization}, Cambridge University Press, 2004.

\bibitem {WLP}
C. Ma, Y. Kamp, L. Willems: \emph{Robust signal selection for linear prediction analysis of voiced speech}, Speech Communication, vol. 12, no. 1, pp. 69--81, 1993.

\bibitem {Alku-WLP}
C. Magi, J. Pohjalainen, T. Backstrom, P. Alku: \emph{Stabilized weighted linear prediction}, Speech Communication, vol. 51, no. 5, pp. 401--411, 2009.

\bibitem {Vahid}
V. Khanagha, K. Daoudi: \emph{An efficient solution to sparse linear prediction analysis of speech},  EURASIP Journal on Audio, Speech, and Music Processing, vol. 3, 2013.


\bibitem {Gardner}
W. Gardner, B. Rao: \emph{Noncausal All-pole Modeling of Voiced Speech}, IEEE Trans. on Audio and Speech Processing, vol. 5, issue 1, pp. 1--10, 1997.

\bibitem {Drugman-Robustness}
T. Drugman, B. Bozkurt, T. Dutoit: \emph{A Comparative Study of Glottal Source Estimation Techniques}, Computer Speech \& Language, Elsevier, vol. 26, issue 1, pp. 20--34, 2012.

\bibitem {Chirp}
T. Drugman, B. Bozkurt, T. Dutoit: \emph{Chirp decomposition of speech signals for glottal source estimation}, ISCA Workshop on Non-linear Speech Processing, 2009.

\bibitem {GCI}
T. Drugman, M. Thomas, J. Gudnason, P. Naylor, T. Dutoit: \emph{Detection of Glottal Closure Instants from Speech Signals: a Quantitative Review}, IEEE Trans. on Audio, Speech and Language Processing, vol. 20, Issue 3, pp. 994--1006, 2012.


\bibitem {GlottalFormant}
B. Bozkurt, B. Doval, C. d'Alessandro, T. Dutoit: \emph{A method for glottal formant frequency estimation}, Proc. ICLSP, 2004.

\bibitem {MGC}
K. Tokuda, T. Kobayashi, T. Masuko, S. Imai: \emph{Mel-generalized cepstral analysis - a unified approach to speech spectral estimation}, Proc. ICLSP, 1994.

\bibitem {TIMIT}
J. Garofolo, L. Lamel, W. Fisher, J. Fiscus, D. Pallett, N. Dahlgren, V. Zue: \emph{DARPA TIMIT Acoustic-Phonetic Continuous Speech Corpus. Tech. rep.}, U.S. Dept. of Commerce, NIST., 1993.

\bibitem{Gini}
S. Rickard, M. Fallon: \emph{The Gini index of speech}, Conf. on Information Sciences and Systems, 2004.

\bibitem {Kurtosis}
N. Hurley, S. Rickard: \emph{Comparing measures of sparsity}, IEEE Trans. Inf. Theory, vol. 55, pp. 4723--4740, 2009.

\bibitem{SRH}
T.Drugman, A.Alwan: \emph{Joint Robust Voicing Detection and Pitch Estimation Based on Residual Harmonics}, Proc. Interspeech, 2011.

\bibitem {SEDREAMS}
T. Drugman, T. Dutoit: \emph{Glottal Closure and Opening Instant Detection from Speech Signals}, Proc. Interspeech, 2009.

\bibitem {Magic}
E. Candes, J. Romberg: \emph{L1-MAGIC: Recovery of sparse signals via convex programming},  California Institute of Technology, Pasadena, 2005.

\bibitem{Hoyer}
P. Hoyer: \emph{Non-negative Matrix Factorization with Sparseness Constraints}, Journal of Machine Learning Research, vol. 5, pp. 1457--1469, 2004.


\bibitem {RESKEW}
T. Drugman: \emph{Residual Excitation Skewness for Automatic Speech Polarity Detection}, IEEE Signal Processing Letters, vol. 20, issue 4, pp. 387--390, 2013.

\bibitem {DSM}
T. Drugman, T. Dutoit: \emph{The Deterministic plus Stochastic Model of the Residual Signal and its Applications}, IEEE Transactions on Audio, Speech and Language Processing, vol. 20, Issue 3, pp. 968--981, 2012.

\bibitem {PCA}
I. Jolliffe: \emph{Principal Component Analysis}, Springer Series in Statistics, 2005.


\bibitem {ARCTIC}
J. Kominek, A. Black: \emph{The CMU Arctic Speech Databases}, SSW5, pp. 223--224, 2004.

\bibitem {GFmodels}
G. Fant, J. Liljencrants, Q. Lin: \emph{A four parameter model of glottal flow}, STL-QPSR4, pp. 1-13, 1985.



%\bibitem {SourceModels}
%G. Fant, J. Liljencrants, Q. Lin: \emph{A four parameter model of glottal flow}, STL-QPSR4, pp. 1-13, 1985.


%\bibitem{Drugman-Thesis}
%T. Drugman : \emph{Advances in Glottal Analysis and its Applications}, PhD Thesis, University of Mons, 2011.



\end{thebibliography}
\end{document}